\documentstyle[prl,aps]{revtex}
\include{epsf}

\begin{document}

%\onecolumn
%\twocolumn[
\title{Motion of a condensate in a shaken and vibrating harmonic trap}
\author{Y.\ Japha$^{\,1,2,3}$ and Y.\ B.\ Band$^{\,2}$}
\address{
$^{\,1}$ Clarendon Laboratory, Department of Physics, University of Oxford, 
Parks Road, Oxford OX1 3PU, UK \\
$^{\,2}$ Department of Chemistry, Ben-Gurion University of the Negev, 
Beer-Sheva, Israel \, 84105 \\
$^{\,3}$ Present Address, LaserComm R\&D, 21 Habarzel St., Tel-Aviv,
Israel \, 61580 }

\maketitle 

\begin{abstract}
%\widetext
The dynamics of a Bose-Einstein condensate (BEC) in a time-dependent
harmonic trapping potential is determined for arbitrary variations of
the position of the center of the trap and its frequencies.  The
dynamics of the BEC wavepacket is soliton-like.  The motion of the
center of the wavepacket, and the spatially and temporally dependent
phase (which affects the coherence properties of the BEC) multiplying
the soliton-like part of the wavepacket, are analytically determined.
\end{abstract}
%\vskip 0.8truecm
%] \narrowtext


Bose-Einstein condensates of dilute atomic gases in magnetic traps
provide a simple many-body system in which to investigate the
evolution of a macroscopic coherent quantum system under the influence
of external forces \cite{BEC-experimental-refs}.  Analytic mean-field
solutions for these systems exist even for time dependent external
forces \cite{Morgan,Dalfovo}.  The special scaling properties of the
harmonic potential, created by the interaction of the atomic magnetic
moments and the average magnetic field of the trap which has a
quadratic spatial form, makes it easy to determine the evolution of
the condensate even under temporal variations of the frequency of the
trap \cite{Kagan,Castin}.  Moreover, Kohn \cite{Kohn} and Dobson
\cite{Dobson} have shown that, for any many-body system in an
arbitrarily changing harmonic potential, the motion of the
center-of-mass of the system is decoupled from the motion of other
degrees of freedom of the system.

Here we present an exact solution for the motion of a Bose-Einstein
condensate under the influence of a harmonic magnetic field whose
center moves as an arbitrary function of time and whose frequency
varies arbitrarily with time.  When the frequency of the harmonic trap
is constant in time, the motion of the condensate is as a rigid body
whose shape is not changed as the potential moves, the motion of the
center of the condensate is analytically determined, and the time
dependence of the phase of the condensate is also analytically
obtained.  We stress that this result applies not only at the
mean-field level of approximation, as described by the
Gross-Pitaevskii equation \cite{Dalfovo}, but quite generally at the
field-theory level.  Hence, upon shaking a harmonic potential, no
matter how vigorously or quickly, a condensate does not develop an
above-the-mean-field component, and does not develop a temperature. 
No amount of shaking will yield a thermal cloud in such a system. 
When the frequency of the harmonic potential also varies, the center
of mass motion of the condensate and its phase can still be
analytically determined, and the condensate shape is obtained by
solving the Gross-Pitaevskii equation for a harmonic potential whose
center is not moving.

We consider a general system of N mutually interacting identical
Bosonic particles of mass $m$ in an external time-dependent harmonic
potential.  The Hamiltonian for the system is given by
\begin{equation}
H = \sum_\alpha\left[\frac{{\bf p}_\alpha^2}{2m}+\frac{1}{2}m
\sum_{i=x,y,z} \omega_{i}^2(t) (x_{\alpha i} - x_{0 i}(t))^2 \right]
+ \sum_{\alpha, \beta \neq \alpha} U({\bf x}_\beta-{\bf x}_\alpha) \,.
\label{eq:N_harmonic}
\end{equation}
Here $x_{0 i}(t)$ is the $i$th component of the time-dependent
position vector of the center of the harmonic trap and $x_{\alpha i}$
is the $i$th component of the $\alpha$th Boson.  Let us make the
coordinate transformation to a new system of coordinates comprising
${\bf X}=\sum_\alpha {\bf x}_\alpha/N$ and ${\bf q}_\alpha = {\bf
x}_\alpha-{\bf X}$ for $\alpha=1,...,N-1$.  Note that ${\bf q}_N={\bf
x}_N-{\bf X}$ is a dependent variable equal to $-\sum_{\alpha=1}^{N-1}
{\bf q}_\alpha$, and $\sum_{\alpha=1}^N {\bf q}_\alpha=0$.  We can
write the quadratic term of the harmonic potential using these
variables as:
\begin{equation}
\frac{1}{2}m \sum_{i=x,y,z} \omega_{i}^2(t) [\sum_\alpha q_{\alpha
i}+X_i-x_{0 i} (t)]^2 = \sum_{i=x,y,z} \omega_{i}^2(t) \left(
\frac{1}{2} M [X_i-x_{0 i}(t)]^2 + \frac{1}{2} m \sum_\alpha q_{\alpha
i}^2 \right) \,,
\label{eq:N_harmonic'}
\end{equation}
where we have defined $M=Nm$ and we used the relation
$\sum_{\alpha=1}^N q_{\alpha i} = 0$.  If, for simplicity of notation,
we drop the indices $i$ that specify the different components of the
harmonic frequency and position variables, the above Hamiltonian now
reads
\begin{equation}
H=[P^2/2M+\frac{1}{2}M\omega^2(t)(X-x_0(t)^2]+\sum_{\alpha=1}^{N-1}
\left[\frac{p_\alpha^2}{2m}+\frac{1}{2}m\omega^2(t)
q_\alpha^2+\sum_{\beta \neq \alpha} U(q_\beta-q_\alpha)\right] \,.
\label{eq:H}
\end{equation}
Hence, the general wavefunction $\psi(x_1,x_2, ...,x_N,t)$ may be
written in product form, $\psi(x_1,x_2, ...,x_N,t) = \psi_{CM}(X,t)
\psi_{rel}(q_1,q_2,...,q_{N-1})$, where the relative part of the wave
function does not depend explicitly on time, the effect of the motion
(time dependence) of the harmonic potential is only on the center of
mass part of the wave function, and this dependence is given via the
the quantity ${\bf x}_0(t)$.  Thus, the motion relative to the center
of mass is decoupled from the motion of the center of mass, and only
the latter is influenced by the shaking.  This is true at the
field-theory level.

At the mean-field theory level, the harmonic potential appearing in
the Gross-Pitaevskii equation is of the form
\begin{equation}
V({\bf x},t)=\frac{m}{2}\sum_{i=x,y,z}\omega_i(t)^2 [x_i-x_{i0}(t)]^2 
\equiv V_0({\bf x}-{\bf x}_0(t),t) \ , \label{harmonic}
\end{equation}
where ${\bf x}_0(t)=(x_0(t),y_0(t),z_0(t))$ is an arbitrary
time-dependent displacement of the center of the trap and
$(\omega_x(t),\omega_y(t),\omega_z(t))$ are the (perhaps
time-dependent) trap frequencies.  Under the action of a potential of
the form (\ref{harmonic}), the internal dynamics of a system of
particles is not affected by an arbitrary motion of the center of the
potential.  This follows from the fact that a quadratic potential of
the form~(\ref{harmonic}) can be expanded at any instant of time
around the mean value ${\bf R}(t)\equiv (R_x(t),R_y(t),R_z(t))$ of the
center-of-mass of the system as
\begin{equation}
V({\bf x},t)=V_0({\bf x}-{\bf R})+m\sum_i\omega_i^2(R_i-x_{i0})x_i 
+\frac{m}{2}\sum_i\omega_i^2(x_{i0}^2-R_i^2) \ . \label{V-expand} 
\end{equation}
The second term in Eq.~(\ref{V-expand}) corresponds to a homogeneous 
time-dependent force acting on the system of particles and it can 
therefore be responsible only for a global shift of the wavefunction 
in position and momentum space.  The third term is 
coordinate-independent and can therefore be responsible only for a 
global phase shift of the wavefunction.

We demonstrate this general principle by considering a condensate of 
alkali atoms in a time-dependent harmonic magnetic trap as given by 
Eq.~(\ref{harmonic}); we produce the general form for the condensate 
wavefunction.  The following is thereby a generalization of the result of
Heller\cite{Heller} for time-dependent harmonic potentials to the case of 
interacting many-body boson systems.  For simplicity we assume that 
the condensate can be described by a single mean-field wavefunction 
$\psi({\bf x},t)$, but this assumption is not necessary because the 
following treatment can be automatically applied if $\psi({\bf x},t)$ is 
regarded as a field operator.  In mean-field, the dynamics of a 
condensate of weakly interacting atoms is determined by the 
time-dependent Gross-Pitaevskii equation,
\begin{equation}
i\hbar\frac{\partial\psi}{\partial t} = 
\left[-\frac{\hbar^2\nabla^2}{2m}+V({\bf x},t)+ 
U_0|\psi({\bf x},t)|^2\right]\psi({\bf x},t) \ , \label{eq:GPE}
\end{equation}
where $U_0 = \frac{4\pi \hbar ^{2}}{m}N\,a_{0}$ is proportional to the 
number of atoms in the condensate and $a_{0}$ is the $s$-wave 
scattering length for collisions between the atoms.  Under the 
influence of the potential of the form~(\ref{harmonic}), the solution 
of this equation can be written as
\begin{equation}
\psi({\bf x},t)=\psi_0({\bf x}-{\bf R}(t),t)\exp\left\{i[{\bf
P}(t)\cdot {\bf x} /\hbar -\phi(t)]\right\} \ ,
\label{psixt}
\end{equation}
where $\psi_0({\bf x},t)$ satisfies the time-dependent Gross-Pitaevskii 
equations with ${\bf x}_0(t)=0$, and the $i$th component of the vector ${\bf 
R}(t)$ satisfies the equation of motion
\begin{equation}
\ddot{R}_i+(\omega_i(t))^2(R_i-x_{i0}(t))=0 \ . \label{eq:Xt} 
\end{equation}
In Eq.~(\ref{psixt}) the momentum ${\bf P}(t) = m\dot{{\bf R}}(t)$, and 
the phase $\phi(t) = \sum_i \phi_i(t)$ is given by 
\begin{equation}
\phi(t)= \frac{m}{2 \hbar} \sum_i \left\{ \int_0^t 
\omega_i^2(t')[R_i^2(t')-x_{i0}^2(t')]+[\dot{R}_i(t')]^2  dt' \right\} \ .
\label{phit} 
\end{equation}

The proof is straight-forward.  Assume that $\psi_0({\bf x},t)$ satisfies 
Eq.~(\ref{eq:GPE}) with ${\bf x}_0 = 0$, then substitute the 
solution (\ref{psixt}) into Eq.~(\ref{eq:GPE}) for arbitrarily varying 
${\bf x}_0(t)$ to obtain
\begin{equation} 
-m\ddot{{\bf R}}\cdot{\bf x} + \hbar \dot{\phi}= 
\frac{m}{2} \dot{{\bf R}}^2 + \frac{1}{2}m\sum_{i}\omega_{i}^2 
[(x_i-x_{i0})^2-(x_i-R_i)^2]\psi_0 \ .
\label{eq:GPE1}
\end{equation}
It is easy to verify that this is indeed satisfied for $R_i(t)$ and 
$\phi(t)$ given above.

Note that the phase factor $\exp\left\{i[{\bf P}(t)\cdot {\bf x} 
/\hbar -\phi(t)]\right\}$ in Eq.~(\ref{psixt}) affects the coherence 
properties of the condensate wavepacket, which are given in terms of 
the coherence function $C({\bf \rho},\tau;t) = \int d^3x \,
\psi^{*}({\bf x}+{\bf \rho},t+\tau) \psi({\bf x},t)$ 
\cite{Tripp_contrast,Hagley}.

As an example, let us first consider the solution in the case where 
$\omega_x(t)=\omega_x={\rm const}$ and determine $R_x(t)$ and $\phi_x(t)$.  
In this case $\psi_0({\bf x})$ may be taken as the ground-state solution of 
the time-independent Gross-Pitaevskii equation, which is obtained by 
replacing the time derivative on the left-hand side of 
Eq.~(\ref{eq:GPE}) by $\mu\psi_0$, where $\mu$ is the chemical 
potential of the condensate.  In this case the solution~(\ref{psixt}) 
is a solitary solution, namely, the condensate moves as a rigid body 
without changing its shape.  The general solution for $R_x(t)$ in 
Eq.~(\ref{eq:Xt}) is then given by
\begin{equation}
R_x(t)=\cos\omega_x t R_x(0)+\frac{\sin\omega_x t}{\omega_x}\dot{R}_x(0)
+\omega_x \int_0^t dt' \sin\omega_x(t-t')x_0(t') \ . 
\end{equation}
In what follows we give some examples for the dynamics in different 
shaking schemes.
For a periodically shaken trap, such that $x_0(t)=x_0\sin\omega t$ and
$R_x(0)=0$,$\dot{R}_x(0)=0$ we obtain the following solution
\begin{equation}
R_x(t) = \frac{\omega_x}{\omega_x^2-\omega^2}\left[\omega_x\sin\omega
t-\omega\sin\omega_x t\right] x_0 \ , \label{sol:per}
\end{equation}
so the instantaneous difference of the center of the wavepacket from 
the center of the potential is given by
\begin{equation}
R_x(t)-x_0(t) = \frac{\omega}{\omega_x^2-\omega^2}\left[\omega\sin\omega
t-\omega_x\sin\omega_x t\right] x_0 \ .
\end{equation}
The expression for the phase $\phi_x(t)$ can be easily obtained 
analytically in terms of simple trigonometric functions using 
Eq.~(\ref{phit}).

We identify three different regimes for this solution:
\begin{enumerate}
\item {\bf The adiabatic regime:} When $\omega\ll \omega_x$, the motion of 
the condensate will adiabatically follow the motion of the center of 
the trap.  We can then approximate
\begin{equation}
R_x(t)-x_0(t) \approx \frac{\omega}{\omega_x}[\frac{\omega}{\omega_x}
\sin\omega t -\sin\omega_x t]x_0 \ ,
\end{equation}
whose maximal value is $\max\{|R_x(t)-x_0(t)|\}\approx x_0\omega/\omega_x$.  
The adiabatic approximation is justified if this is much smaller than 
the spatial width $\Delta x$ of the condensate wavefunction $\psi_0$, 
i.e., $\omega/\omega_x\ll \Delta x/2 x_0$.

\item {\bf Resonance:} If $\omega=\omega_x$, then 
\begin{equation}
R_x(t)=\frac{x_0}{2}\left[\sin\omega_x t-\omega_x t\cos\omega_x t\right] \ ,
\end{equation}
and the amplitude grows linearly with time. 

\item {\bf Averaged effective potential:} When $\omega\gg \omega_x$,
\begin{equation}
R_x(t) \approx x_0\frac{\omega_x}{\omega}\sin\omega_x t \ , 
\end{equation}
and the motion of the center of the condensate is only slightly 
affected by the shaken trap.

\end{enumerate}

\begin{figure}
\centerline{\epsfxsize=6.25in\epsfbox{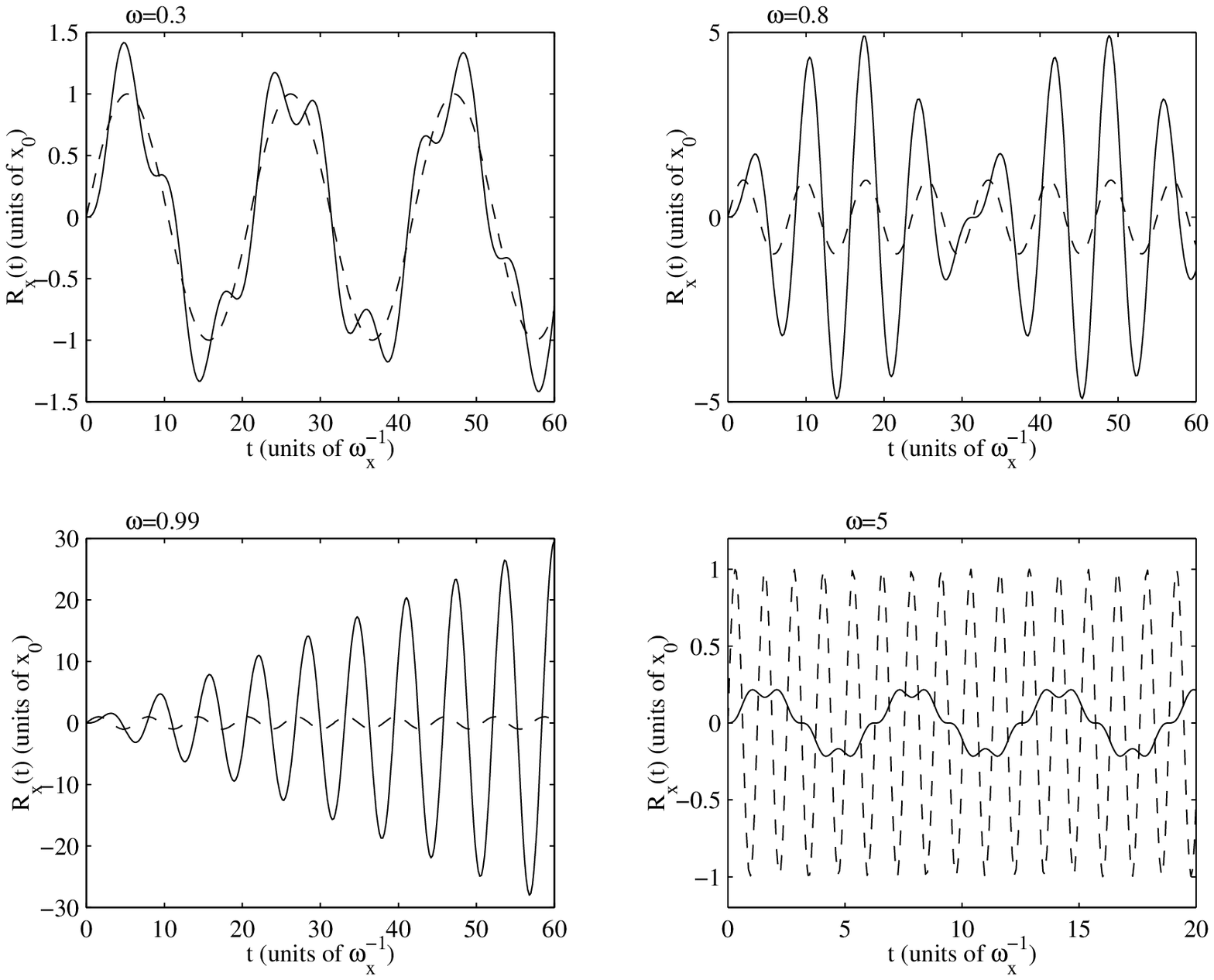}}
%\begin{center}
\caption[f1] {$R_x(t)$ for a periodically shaken trap with constant
trap frequency $\omega_x$ and four different values of the shaking frequency
$\omega$.  The dashed curve is the position of
the center of the trap $x_0(t)$.}
%\end{center}
\label{fig1}
\end{figure}

Fig.~\ref{fig1} plots $R_x(t)$ versus $t$ for four different values 
of the ratio $\omega/\omega_x$.  In Fig.~\ref{fig1}a, $\omega/\omega_x = 0.3$ 
and the small deviation of $R_x(t)$ from $x_0(t)$ is evident.  
In Fig.~\ref{fig1}b, $\omega/\omega_x = 0.8$ and a substantial overshoot 
of $R_x(t)$ relative to $x_0(t)$ is obtained.  The linear growth of 
$R_x(t)$ is clearly seen for $\omega/\omega_x = 0.99$ in Fig.~\ref{fig1}c, 
and the small oscillation of $R_x(t)$ is evident for $\omega/\omega_x = 5$ in 
Fig.~\ref{fig1}d.

For two dimensional motion corresponding to an ellipsoidal rotation of 
the center of the trap, $x_0(t)=x_0 \sin\omega t$ and $y_0(t)=y_0 
[1-\cos\omega t]$, the solution for $R_x$ is given by Eq.~(\ref{sol:per}) 
and $R_y$ is
\begin{equation}
R_y(t)=\frac{1-\cos\omega_y t}{\omega_y} + 
\frac{\omega_y}{\omega_y^2-\omega^2}[\cos\omega_y t-\cos\omega t] \ .
\end{equation}
The phase $\phi(t)$ is again simple to calculate.

If the trap is periodically shaken in the $x$ direction for a finite 
duration $t_0$ and then the shaking is stopped, then the solution for
the center of the wavepacket for $0<t<t_0$ is given by Eq.~(\ref{sol:per}), 
and for $t>t_0$,
\begin{equation}
R_x(t>t_0) = \frac{\omega_x}{\omega_x^2-\omega^2}x_0
\left\{\left[\omega_x\sin\omega t_0-\omega\sin\omega_x t_0
\right]\cos\omega_x t +\omega\left[\cos\omega t_0-\cos\omega_x
t_0\right]\sin\omega_x t\right\} \ .
\end{equation}

Analytic solutions to Eq.~(\ref{eq:Xt}) for arbitrary $\omega_i(t)$ are 
not known.  Even for harmonically varying $\omega_i(t)$ and $x_{i0}(t)$, 
where Eq.~(\ref{eq:Xt}) corresponds to a driven Mathieu equation 
\cite{AB-Heller2}, analytic solutions are not available.  In this case, 
depending on the ratio of the frequencies for the variation of 
$\omega_i(t)$ and $x_{i0}(t)$, regular bounded motion or unbounded motion 
of $R_i(t)$ may result.  Nevertheless, Eq.~(\ref{psixt}) with central 
wavepacket coordinate ${\bf R}(t)$ and phases $\phi_i(t)$ given by 
Eqs.~(\ref{eq:Xt}) (\ref{phit}) gives the analytic form for the 
wavefunction.

\begin{figure}
\centerline{\epsfxsize=6.25in\epsfbox{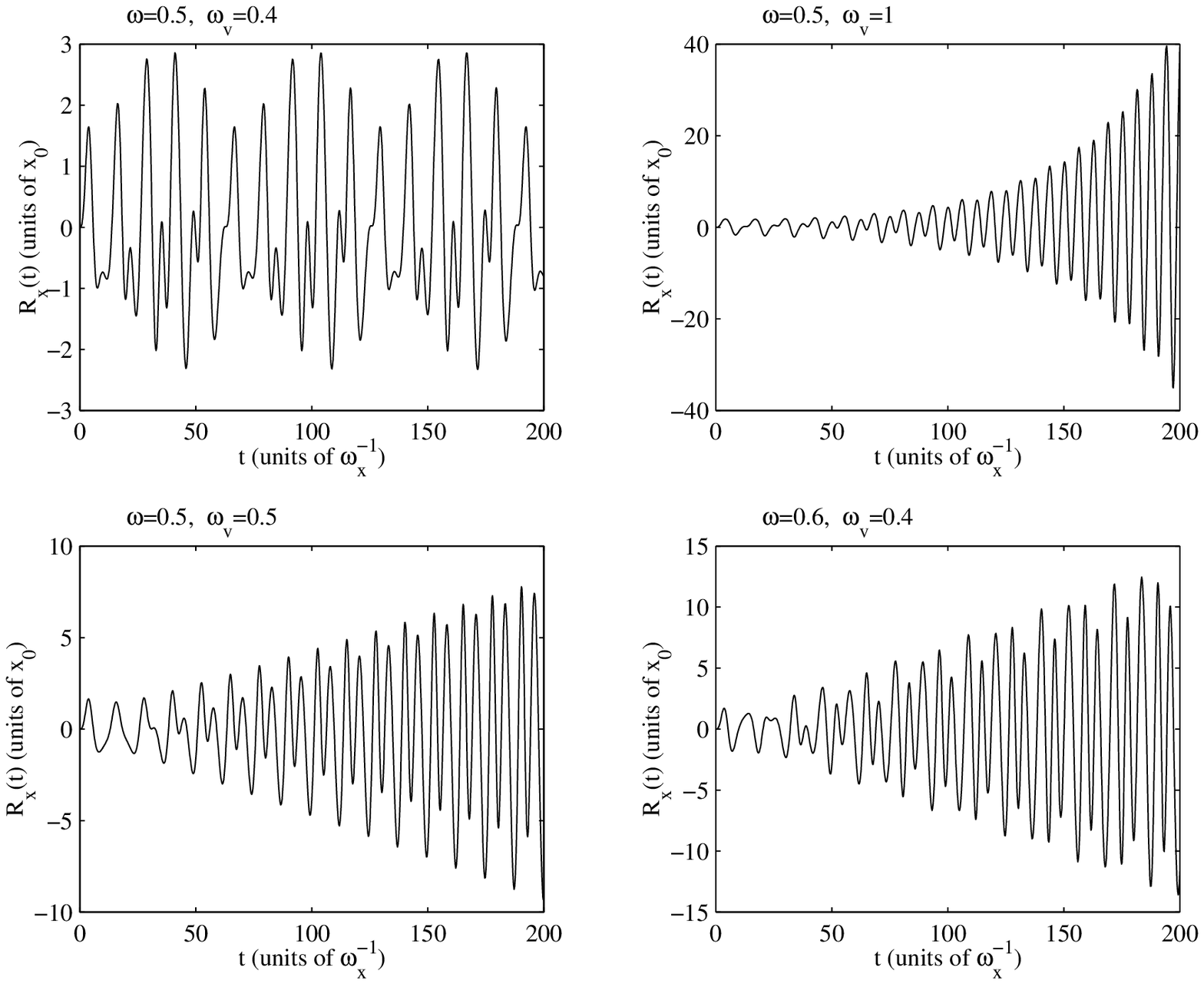}}
%\begin{center}
\caption[f2] {$R_x(t)$ for a periodically shaken trap with periodically
vibrating trap frequency $\omega_x(t)=\omega_x(0)[1+0.25\sin(\omega_v t)]$ for
different values of the shaking frequency $\omega$ and vibration frequency
$\omega_v$. }
%\end{center}
\label{fig2}
\end{figure}

The following numerical example illustrates the nature of the analytic 
solution for a shaken vibrating trap.  Fig.~\ref{fig2} plots $R_x(t)$ 
versus $t$ for a periodically varying trap frequency 
$\omega_x(t)=\omega_x(0)[1+\delta\sin(\omega_v t)]$.  where $\delta$ and 
$\omega_v$ are the amplitude and frequency of the vibration.  Unbounded 
motion is expected when either the shaking or the vibration 
frequencies are resonant with the basic trap frequency $\omega_x(0)$ 
(Fig.~\ref{fig2}b), or if the sum of the vibration frequency and shaking 
frequency, or the sum of integers times these frequencies, is resonant with 
$\omega_x(0)$ (Fig.~\ref{fig2}c,d).  Unbounded solutions for the volume 
of the condensate are expected even in the absence of shaking when the 
vibration frequency is resonant with the trap 
frequency \cite{Kagan,Castin}.  However, the motion of the center of 
the trap causes the center of mass of the condensate to move with 
respect to the center of the trap, and the amplitude of this motion 
may be amplified by a resonant change of the trapping frequency.

We stress that the main result of this paper is valid for any system
of interacting particles in a harmonic potential. Multi-component BECs
with harmonic potentials have solutions which also have the properties
discussed above, provided that the time-dependent harmonic potentials
for the various components are exactly the same. Since the magnetic
moments of atoms in different Zeeman levels, or different isotopic
species of the same element, or of different elements (for mixed
species BECs) are in general not identical, our solution will in
general not be relevant for these cases.

In practice, magnetic traps for BEC alkali atoms are harmonic only 
near their center within a range $r_h$, which is typically a few tens 
of microns.  If such a magnetic trap is shaken, the condensate may 
enter a region where the true potential is anharmonic.  In this case 
the shape of the condensate may change and the motion of its center of 
mass may deviate from the solutions given above.  The maximal velocity 
that can be achieved by accelerating a condensate within the harmonic 
range $r_h$ is roughly given by $v_{max}=\omega_{trap}r_h$, which is 
typically of the order of a few cm/sec.  By shaking a magnetic trap 
using a time-dependent gradient magnetic field it is possible to boost 
condensates in a controlled manner to this range of velocities without 
changing their shape.  This method can also be used in conjunction with
other methods of optically output coupling high momentum wavepackets
\cite{nist_oc,mit_bragg} to create novel wavepackets.

To summarize, quite generally, shaking a a harmonic trap will not
cause a thermal cloud of atoms to develop from a condensate state;
only the center of mass motion of the condensate is affected.  This is
true even at the field theory level.  We determined analytic solutions
for the dynamics of Bose-Einstein condensates of dilute atomic gases
in shaken and vibrating harmonic traps as described by the
Gross-Pitaevskii equation (the mean-field level of approximation). 
One potential application of relevance for the field of atom optics is
to boost BECs to desired velocities.

\bigskip

Y.J. acknowledges financial support from EC-TMR grants.  This work was
supported in part by grants from the US-Israel Binational Science
Foundation (grant No.  98-421), Jerusalem, Israel, the Israel Science
Foundation (grant No.  212/01), the Israel MOD Research and Technology
Unit, and the National Science Foundation through a grant for the
Institute for Theoretical Atomic and Molecular Physics at Harvard
University and Smithsonian Astrophysical Observatory (Y.B.B.).

\newpage

\bigskip

\begin{thebibliography}{99}

\bibitem{BEC-experimental-refs}
Stamper-Kurn, D. M., H.-J. Miesner, S. Inouye, M. R. Andrews,
and W. Ketterle, Phys. Rev. Lett. {\bf 81}, 500 (1998);
Miesner, H.-J., D. M. Stamper-Kurn, M. R. Andrews, D. S.
Durfee, S. Inouye, and W. Ketterle, Science {\bf 279}, 1005 (1998);
Inouye, S., M. R. Andrews, J. Stenger, H.-J. Miesner, D. M.
Stamper-Kurn, and W. Ketterle, Nature (London) {\bf 392},
151 (1998);
Jin, D. S., J. R. Ensher, M. R. Matthews, C. E. Wieman, and E.
A. Cornell, Phys. Rev. Lett. {\bf 77}, 420 (1996);
Jin, D. S., M. R. Matthews, J. R. Ensher, C. E. Wieman, and E.
A. Cornell, Phys. Rev. Lett. {\bf 78}, 764 (1997);
Mewes, M.-O., M. R. Andrews, N. J. van Druten, D. M. Kurn,
D. S. Durfee, C. G. Townsend, and W. Ketterle, Phys.
Rev. Lett. {\bf 77}, 988 (1996).

\bibitem{Morgan} S. A. Morgan, R. J. Ballagh and K. Burnett, Phys. 
Rev. {\bf A 53}, 4338 (1997).

\bibitem{Dalfovo}  F. Dalfovo {\it et al.}, Rev. Mod. Phys. {\bf 71}, 463
(1999).

\bibitem{Kagan} Yu. Kagan, E.L. Surkov, and G.V. Shlyapnikov, Phys. Rev. 
{\bf A 54}, R1753 (1996); {\bf 55}, R18 (1997).

\bibitem{Castin} Y. Castin and R. Dum, Phys. Rev. Lett. {\bf 77}, 5315 
(1996); {\bf 79}, 3553 (1997).

\bibitem{Kohn}
W. Kohn, Phys. Rev. {\bf 143}, 1242 (1961).

\bibitem{Dobson} J. F. Dobson, Phys. Rev. Lett. {\bf 73}, 2244 (1994).

\bibitem{Heller}
E. J. Heller, J. Chem. Phys. {\bf 62}, 1544 (1975).

\bibitem{Tripp_contrast} 
M. Trippenbach, Y. B. Band, M. Edwards, M. Doery, and P. S. Julienne, 
% ``Coherence properties of an atom laser'',
J. Phys. {\bf B33}, 47 (2000).

\bibitem{Hagley}
E. W. Hagley, L. Deng, M. Kozuma, M. Trippenbach, Y.B. Band, M.
Edwards, M. Doery, P.S. Julienne, K. Helmerson, S.L. Rolston, and
W.D. Phillips, Phys. Rev. Lett. {\bf 83}, 3112 (1999).

\bibitem{AB-Heller2}
M. Abramowitz and I. A. Stegun, {\it Handbook of Mathematical 
Functions}, (Dover, NY, 1972).  See also,
E.J. Heller, J.L. Ozment and  D.W. Pratt,
J. Chem. Phys. {\bf 88}, 2169 (1988).

\bibitem{nist_oc}
M. Kozuma, L. Deng, E. Hagley, J. Wen, R. Lutwak, K. Helmerson, S.L.
Rolston, and W.D. Phillips, Phys. Rev. Lett.{\bf 82}, 871 (1999).

\bibitem{mit_bragg}
J. Stenger, S. Inouye, A.P. Chikkatur, D.M. Stamper--Kurn, D.E.
Pritchard, and W. Ketterle, Phys. Rev. Lett.{\bf 82}, 4569 (1999).

\end{thebibliography}
\end{document}